# Bose's derivation, the crucial link to quantum mechanics

Urjit A. Yajnik

*Indian Institute of Technology Bombay*

## 1. Introduction

It is a pleasant duty to pay tributes to a great scientist, originator a pioneering work coming from India during a period when the centre of the intense activity was Europe. In this brief note I address the question not frequently asked, namely, why did it take two decades between Einstein's first proposal of photons and derivation of the full Planck formula from first principles of Statistical Mechanics, albeit with a fundamental new approach to counting of states. Secondly, we acknowledge that British Calcutta was very much an active centre of science and the humanities. Yet the ecosystem of research science was nowhere as developed as in Europe. Thus the question, why did it fall to an independent inquirer in far away Dacca to arrive at the correct derivation of this formula, arguably a most crucial one of the first half of the twentieth century? I also argue that the timing of Bose's communication to Einstein played a crucial role in the emergence of the definitive version of quantum mechanics in the late 1920's.

## 2. A quarrel with the stalwarts

The year 1905 is considered the miraculous one for Einstein. The five papers (one of them his doctoral thesis) seemed to deal with three rather independent areas of research (1) atomistic view of matter and Brownian motion (2) Special Relativity as the correct generalisation of Galilei group of space time transformations arising out of Maxwell's equations (3) the lonesome paper on the much misunderstood "light quantum" or the photon hypothesis. But a study of available sources on Einstein's thought processes and personal communications during this period reveals that underlying all of these was the atomistic view of the fundamental constituents of nature.

Special Relativity is known for its having done away with the ether as propping up electromagnetic waves. In our context it can also be construed to mean that light had an autonomous existence, and carried energy, much like any matter particle. Light was not an "emergent" phenomenon needing a medium to express itself. In a sense the $E=mc^2$ may be understood to be the bridge between radiation and matter, wherein absorption of light causes increase in the rest mass of a matter particle.

This gave rise to the next obvious concern, why is the dynamical description of light (Maxwell's equations) so different from that of "ponderable matter" (Newtonian mechanics and hydrodynamics). Einstein's photon paper may be seen as addressing this precise question. Indeed these are the opening lines of that paper [1]:

*"A profound formal difference exists between the theoretical concepts that physicists have formed about gases and other ponderable bodies, and*

*maxwell's theory of electromagnetic processes in so-called empty space".*

Planck's 1900 formula which had remained a mystery to most physicists provided Einstein with the needed data to attack this problem. Einstein's explicit viewpoint was that in emission and absorption of light it must be considered to be a quantum of energy, "propagating undivided into ever expanding space". Einstein used the ultraviolet end of Planck's formula, effectively Wien's formula to deduce that the volume dependence of the entropy of light seems to scale as if it contains a number of quanta carrying energy specified by Planck's law, E=$h\nu$.

But while using Boltzmann's definition of entropy, S=k *log* W, Einstein has a serious objection to its interpretation as "proportional to the number of complexions of a state of the system" [2]. Einstein felt that this interpretation requires time averaged occupation, or the fraction of time the system spends in that state. In order to not commit to this interpretation Einstein spends time in his photon paper to prove that the formula for entropy is applicable to the case he wants to apply it to, while expressing doubt about the rest of Boltzmann's formalism. He asserts [1],

*"I will show in a separate paper ... and hope to eliminate a logical difficulty that still obstructs the application of Boltzmann's principle."*

This point is noted because it may be a clue to why Einstein himself did not provide the proof that S N Bose provided two decades later.

## 3. A suspect paper

In order to appreciate true significance of Bose's paper we must recapitulate the fate of the photon paper, and along with it that of Einstein's reputation. If everyone understood Einstein's underlying viewpoint they would have been satisfied to note that light quanta are like other ponderable particles, and can take away energy in packets when created and give up energy when absorbed. It would have rounded out the entire atomistic revolution begun with Dalton, Avogadro, Canizzaro and being brought to completion by Boltzmann. But such was not the welcome it received. While Einstein's application of similar principles to vibrational modes of a lattice was welcomed, the photon paper was seen as suspect, not least because it was patently obvious to everyone familiar with Marconi's inventions that light was electromagnetic waves being carried all the way across the Atlantic.

This paper prevented Einstein's acceptance to the Prussian Academy, and in turn created difficulties in him receiving the Nobel Prize. He was under pressure from his well wishers to retract this paper. However in the exclusive and prestigious Solvay conference of 1911, given the opportunity to make this crucial admission, he went on to assert :

*"I insist on the provisional character of this concept, which does not seem reconcilable with the experimentally verified consequences of the wave theory, ... but must assert that it is applicable to the domain of phenomena for which it is proposed."*

Ironically however, the same paper served as the reason for which to give Einstein his Nobel Prize. By 1913 Robert Millikan set out to check the photoelectric effect, hoping to disprove Einstein. This was a difficult experiment, as every point in the graph is a different source which needs calibration. But when completed it actually verified Einstein's formula. Millikan proceeded to write the conclusion to his paper, by first dismissing Einstein's explanation. Probably by misinterpreting Einstein's German language disclaimer made at the Solvay conference, Millikan assumed that Einstein had retracted the explanation himself. Nevertheless, the formula had now been empirically verified. And for the 1922 Nobel Prize, the committee cautiously cited Einstein's 1905 paper for "correct prediction of the photoelectric formula". The prize was not for the physical explanation it provided, nor for its conceptual basis, but only for the correct prediction of the empirically verified law.

A crucial point to be noted is that Einstein never provided the Statistical Mechanics derivation of the Planck distribution formula now routinely taught to undergraduates. In the decades that followed he focussed his profound insights and powers of simplification to a variety of settings that would elucidate the formula heuristically, arriving inter alia at the famous A and B coefficients, and to the pioneering concept of a laser. But that simple derivation that would vary the possible ways of assigning photons to their energy levels, to show that it was extremised by Planck distribution was not forthcoming.

We may wonder, why so? The answer seems to lie in Einstein's misgivings about Boltzmann's method and its assumptions about weightages to be assigned to states. We may conjecture the following two points

1. Much debate was going on regarding Boltzmann's works at the turn of 1900's. Some commentators have noted that Boltzmann had philosophical writing style that was discursive, and the concepts he defined underwent changes in their meaning even within the same long essay. This may have reinforced Einstein's suspicion of Boltzmann's method.

2. J. Willard Gibbs had already published a definitive text that resolved Boltzmann's problems at least in equilibrium statistical mechanics. Next I make a rather far fetched conjecture in the history of science : this opus, published in 1902 by Yale University Press in English language was probably not known to or not accessible to Einstein, also perhaps due to the language. Delays in mutual communication between European societies and the Royal Society are well known. Until republished in translation by the local society, a foreign publication probably did not receive enough attention.

Above considerations make clear the uncanny predicament surrounding the photon hypothesis. *While the whole world, especially the English speaking world, trusted Boltzmann's methods duly refined by Gibbs, they did not believe in photons. Einstein who so stubbornly believed in photons, at the risk to his career and his Nobel, did not trust Boltzmann's methods and did not provide a lucid derivation of Planck's formula starting with the photon hypothesis.*

## 4. A resolution from far away

It appears then that S N Bose an avid student of Einstein and the other contemporary developments in Europe and with his unique brilliance in grasping these developments, was the one well positioned to apply Boltzmann's method to Einstein's photons. Even so, this final stretch of the story is not so simple.

Reading Bose's communication and on pursuing Einstein's correspondence with others, the emphasis seems to be in an ingenious derivation of the front factor, the phase space factor $8\pi v^3/c^3$. The point is that Planck onwards, everyone took this factor as arising from the degeneracy of wave modes that would fit in the cavity. In Bose's paper he proposed dividing the phase space into quantal units of size $h^3$. And he arrived at this front factor as the degeneracy factor for particulate photons filling this discretised phase space. Bose himself seems quite excited by this conceptual advance in his paper. And indeed this is how we teach the derivation of the phase space factor today. It did cause puzzlement to Bose as an extra factor of 2 was needed, later understood as the number of independent helicity states, a concept foreign to classical reasoning. But other than that, in Bose's eyes, this derivation was the crowning glory of the paper.

However, the real innovation in Bose's paper remained obscure for some time. It is that he also relied on Planck's original reasoning while deriving his number of ways of distributing "energy elements" among the oscillators. Plank's derivation which heavily relied on the oscillators in the walls of the cavity, inadvertently counted these "energy elements" as generic quantities, not of autonomous physical significance. In this way what we now refer to as "indistinguishability" had secretly entered Planck's derivation.

When Bose adopted Boltzmann's method to Einstein's photons, he distributed the latter in the same way as Planck's "energy elements". But now the latter had the physical significance of photons, quanta of light capable of being emitted and absorbed and capable of propagating as independent entities in free space.

## 5.  A complementary hypothesis

What appears to be a most amusing coincidence of history is that de Broglie's wave hypothesis emerged at exactly the same time as Bose's derivation. And both these path breaking works landed on Einstein's desk in the year 1924.

During autumn of 1923, de Broglie began to think of Einstein's quanta and proposed to make a complementary assumption about matter particles, namely that they have waves associated with them. He made presentations to the French Academy and submitted his thesis in November [3]. He also predicted that this could be observed in electron diffraction. However the members of the academy were not convinced, but the thesis was not rejected outright, probably because the de Broglie's were a distinguished family. Louis was a duke and his brother had been a president of the Academy. To be on the safe side they sent the thesis to a foreign academy member, Albert Einstein of Prussian Academy. It would be up to him to accept or reject it. Einstein does not seem to have responded immediately.

In July 1924 Bose sent his famous paper to Einstein with a request to communicate it. This must have been the biggest "a-ha" or *eureka* moment for Einstein. The photons he so dearly believed for two decades were now theoretically on the strongest footing as particles. He was thus confident in approving de Broglie's thesis, the complementary hypothesis that electrons also behave as waves. Loius de Broglie defended his thesis in November 1924. Electron diffraction was observed in experiments by G. P. Thomson and by Davisson and Germer in 1927. Victor-Louis de Broglie received Nobel Prize for his hypothesis in 1929.

Einstein considered de Broglie's hypothesis to be " a first feeble ray of light on this worst of our physics enigmas". Bose's derivation does not seem to have elicited any comparable admiration. Einstein proceeded to extend it himself to massive particles without informing Bose, and reached the amazing conclusions about what is now called Bose-Einstein condensation. In retrospect, and as we recapitulate in the conclusion, Bose's derivation was as revolutionary as de Broglie's. But it was too premature to see its full significance at that point, and it did not receive the Nobel Prize. The world does however credit Bose for his outstanding derivation that brought the photon concept to its due fruition, and the term *boson* has become a common nomenclature for all quantum excitations of integer or zero spin.

## 6. Conclusion

It is a rather strong assumption that Einstein did not believe Boltzmann's method due to his having not familiarised himself with Gibb's text. Perhaps there were other considerations why Einstein did not attempt the derivation Bose did. It could also be due to the need for the subtle counting inadvertently introduced by Bose based on Planck's work, without which the derivation would not go through. Be that as it may, it fell to Bose to do everything right, in a way that this uncanny derivation can now be taught to college students.

A very important point seems to elude the public discourse on quantum mechanics, where the wave particle duality gets over emphasised, but the uncanny counting required for the states of a quantum system, which bears on the fundamental concept of the identity of an entity is not emphasised. The situation is illustrated in Figure 1.

This state of affairs has two unfortunate nomenclatures associated with it. One is the reference to "statistical" weight. This creates a sense of uncertainty involved, or indeterminable or superfluous pieces of information that can be ignored or averaged over. This is not the fact. It is simply the unusual counting of states of two quanta, even when we have a small number of them and no averaging or statistics is involved. In this sense quanta are not particles. The number operator exists, and its eigenvalues characterise the possible independent states. But the state with number operator eigenvalue 2 is not simply a state with two photons in the classical sense.

|        | Classical     | Quantum:B-E   | Quantum:F-D |
|--------|---------------|---------------|-------------|
| H H    | $\frac{1}{4}$ | $\frac{1}{3}$ | 0           |
| HT or TH | $\frac{1}{2}$ | $\frac{1}{3}$ | 1         |
| TT     | $\frac{1}{4}$ | $\frac{1}{3}$ | 0           |

Figure 1 : The statistical weightage to be associated with the toss of two identical coins in classical, Bose-Einstein and Fermi-Dirac cases.

Due to the fact that one uses 1-particle states as a convenient basis, one is then led to the illusion that the two photons (or fermions) are "entangled". In reality that state whose Hilbert space weightage is unity, was created like that, not a mysterious mix up between two independent states.

In conclusion, of the several mysteries being touted for the quantum world, the only one that persists and challenges common sense is that outcomes of experiments for otherwise presumed identical systems can be different statistically. However the two principles that need to be taught early enough and not mystified, are the principle of linear superposition of states, providing the convenient scaffolding of a "wave function", and the necessity to set up Bose or Fermi type states out of 1-particle states by symmetrising or anti-symmetrising. This counting should be considered as a foundational postulate of quantum mechanics and taught along with the basics.

Bose was the first to apply the new enumeration of states to the case of thermal distribution of photons, and his communication to Einstein coincided with the latter's deliberations over de Broglie's thesis. Einstein could see the truth of both the proposals and would have felt elated at the beautiful derivation by Bose which treated photons intrinsically as particles, and the perfect complementarity to ponderable matter provided in de Broglie's thesis. Einstein's conception of the atomistic properties of the building blocks of nature as envisaged in his 1905 papers, came to a crowning conclusion with these two papers, and opened up the path to further development of Quantum Mechanics.

## Acknowledgement :

The author wishes to thank Prof. S. P. Pandya for many inspiring conversations.